\documentclass[12pt,preprint]{aastex}

\usepackage{graphicx}
\usepackage{epsfig}
\usepackage{lscape}

\begin{document}
\slugcomment{}

\shorttitle{Discovery of $\gamma$-ray pulsation and X-ray emission from the black widow pulsar PSR J2051-0827}

\title{Discovery of $\gamma$-ray pulsation and X-ray emission from the black widow pulsar PSR J2051-0827}
\shortauthors{Wu et al.}

\author{J. H. K. Wu\altaffilmark{1}, A. K. H. Kong\altaffilmark{1,2}, R. H. H. Huang\altaffilmark{1}, J. Takata\altaffilmark{3}, P. H. T. Tam\altaffilmark{1}, E. M. H. Wu\altaffilmark{3}, K. S. Cheng\altaffilmark{3}}

\altaffiltext{1}
{Institute of Astronomy and Department of Physics, 
National Tsing Hua University, Hsinchu, Taiwan }
\altaffiltext{2}
{Golden Jade Fellow of Kenda Foundation, Taiwan}
\altaffiltext{3}
{Department of Physics, University of Hong Kong, Pokfulam Road, 
Hong Kong}

\email{wuhkjason@gmail.com, akong@phys.nthu.edu.tw}


\begin{abstract}
We report the discovery of pulsed $\gamma$-ray emission and X-ray emission
from the black widow millisecond pulsar PSR J2051-0827 by using the
data from the Large Area Telescope (LAT) on board the Fermi Gamma-ray
Space Telescope and the Advanced CCD Imaging Spectrometer array
(ACIS-S) on the Chandra X-ray Observatory. Using 3 years of LAT data,
PSR J2051-0827 is clearly detected in $\gamma$-ray with a signiÞcance of $\sim$ 8$\sigma$
in the 0.2 - 20 GeV band. The 200 MeV - 20 GeV $\gamma$-ray spectrum of PSR J2051-0827 can be
modeled by a simple power-law with a photon index of 2.46 $\pm$ 0.15.
Significant ($\sim 5\sigma$) $\gamma$-ray pulsations at the radio period were detected. PSR J2051-0827 was also
detected in soft (0.3--7 keV) X-ray with Chandra. By comparing the observed $\gamma$-rays and X-rays with
theoretical models, we suggest that the $\gamma$-ray emission is from the
outer gap while the X-rays can be from intra-binary shock and pulsar
magnetospheric synchrotron emissions.

\end{abstract}


\section{INTRODUCTION}
It is widely accepted that millisecond pulsars (MSPs) have been spun up by accreting masses from a close binary companion (Alpar et al. 1982). A sub-class of binary systems consist of a MSP and a very low mass companion ($M_{c}<0.05M_{sun}$) orbiting in a tight orbit ($P_{b}<24 hr$). In such systems particles or $\gamma$-rays emitted from the pulsar are thought to be ablating the companion, which would eventually destroy the companion, resulting in a solitary MSP (Fruchter et al. 1995; Takata et al. 2011); MSPs in this kind of binary system are usually called Black Widow pulsars.

PSR J2051-0827 is the second black widow pulsar after PSR B1957+20 (Fruchter, Stinebring $\&$ Taylor 1988). PSR J2051-0827 was discovered in 1996 in the Parkes all-sky survey of the southern sky for low-luminosity and millisecond pulsars (Stappers et al. 1996). The spin period of PSR J2051-0827 is 4.5ms. The MSP is orbiting around a very low mass companion ($m_{c}<0.027 M_{sun}$) with a orbital period of 2.4 hrs, thus joining the black widow populations (Stappers et al. 1996). Since the discovery of this pulsar, follow-up observations have been performed to reveal the nature of this black widow system (Stappers, Bessell $\&$ Bailes 1996; Stappers wt al. 1999; Stappers et al. 2001b; Doroshenko et al. 2001; Lazaridis et al. 2011). 

Before the launch of the {\itshape Fermi}-LAT, all black widow pulsars have been found by radio surveys but most of them are in globular clusters (GCs). As GCs usually hold a large number of MSPs and many GCs are $\gamma$-ray emitters (Abdo et al. 2010; Kong et al. 2010; Tam et al. 2011). However,  we cannot resolve individual pulsars in $\gamma$-rays in GCs due to the limited angular resolution of {\itshape Fermi}-LAT, which makes studies of $\gamma$-ray black widow pulsars in GCs challenging. After the launch of the {\itshape Fermi}-LAT, more and more field black widows were identified as {\itshape Fermi} sources, opening a new observation window of Black Widow pulsars studies, providing a larger sample for characteristic studies on black widow pulsars. So far a total of 10 black widow pulsars have been found in the Galactic field, in which 7 of them are Fermi sources and at least 3 of them show pulsed $\gamma$-rays (Roberts 2011). Among those which has remained undetectable in $\gamma$-rays, PSR J2051-0827 is the closest one with a $d_{NE2001}$~distances of $1.04$~kpc (Lazaridis et al. 2011),  which increases the chance of detecting it in $\gamma$-rays despise it has a relativity low spin down power $\dot{E}=0.53\times10^{34}~erg/s$ (Stappers et al. 1996). In this letter, we present the discovery in $\gamma$-ray and X-ray using {\itshape Fermi}-LAT and {\itshape Chandra} data. 

\section{DATA ANALYSIS AND RESULTS}
\subsection{GAMMA-RAY DATA}
In our analysis, we used the LAT data between 2008 August and 2011 September ($\sim$3 years of data). 
To reduce and analyze the data, the {\itshape Fermi} Science Tools v9r23p1 package, which is 
available from the {\itshape Fermi} Science Support Center\footnote{
http://fermi.gsfc.nasa.gov/ssc/data/analysis/software/}, was used. We used Pass 7 data and  
selected events in the ``Diffuse" class (i.e.~event class 2) only. 
In addition, we excluded the events with zenith angles larger than 100$\degr$ to 
reduce the contamination by Earth albedo gamma-rays. The instrumental response functions (IRFs) ``P7SOURCE\_V6" were adopted throughout this study.

\begin{figure}[t]
\centerline{\psfig{figure=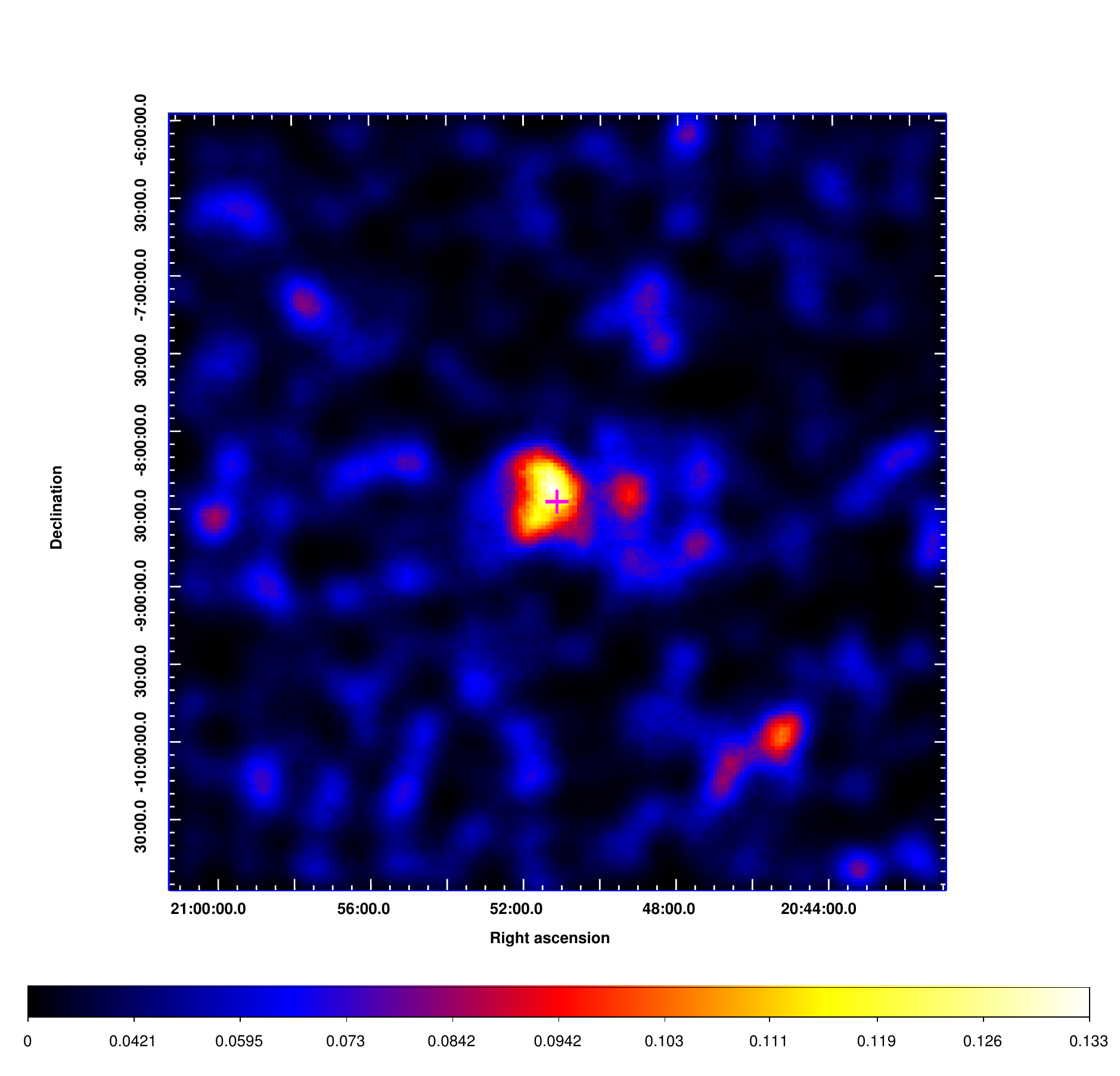,width=8cm,clip=,angle=0}}
\caption[]{2\degr$\times$2\degr~LAT count map in 1-20 GeV centered at PSR J2051-0827 with 
a pixel size of 0.025\degr, smoothed by a Gaussian kernel of 0.15\degr. 
The cross in magenta represents the nominal position of PSR J2051-0827. A scale bar is used to indicate counts/pixel. No background subtraction 
has been applied.}
\label{cmap_psra}
\end{figure}

Events were selected within a circular region-of-interest (ROI) 
centered at the nominal position of PSR J2051-0827 ~(i.e. RA=$20^{h}$$51^{m}$$07^{s}.51$ Dec=$-08^{\circ}27^{\arcmin}37^{\arcsec}.7$).
In order to reduce systematic uncertainties and achieve a better background modeling, a circular ROI with a diameter of $10\degr$ was adopted throughout the unbinned likelihood analysis. For a preliminary inspection of the chosen part of the sky around PSR J2051-0827, a binned photon count map in $1-20$~GeV was produced with task \textit{gtbin} (see Figure~\ref{cmap_psra}). $\gamma-$ray excess can be clearly identified around the nominal position of PSR J2051-0827 even
before subtracting the background. 

\begin{figure}[t]
\centerline{\psfig{figure=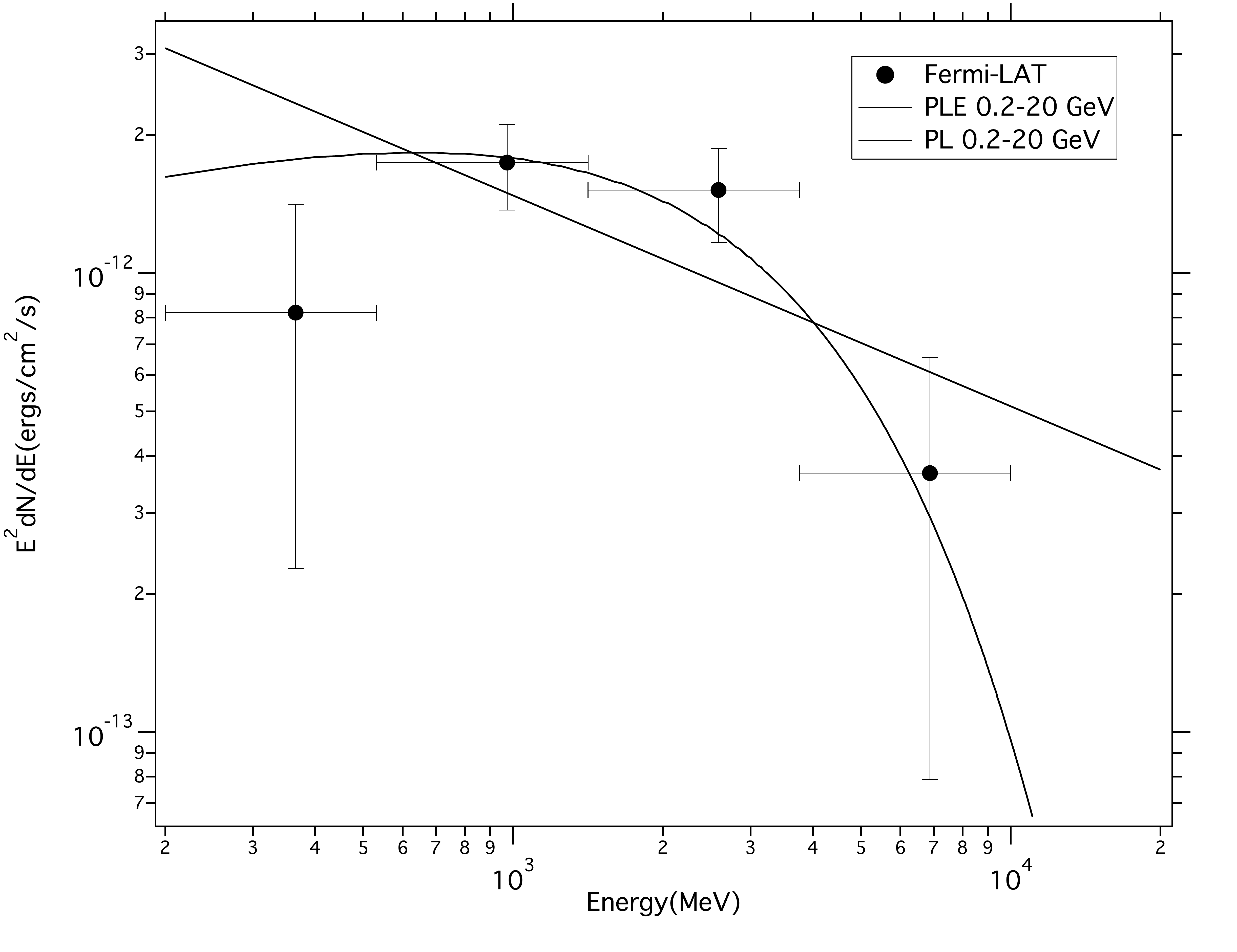,width=8cm,clip=,angle=0}}
\caption[]{\emph{Fermi} LAT spectrum of PSR J2051-0827 in the $0.2-10$~GeV band. Curve and straight line represent the best-fit PLE \& PL model respectively.}
\label{lat_spec}
\end{figure}

To investigate the spectral characteristics of PSR J2051-0827, we performed an unbinnned likelihood analysis with \textit{gtlike}, by assuming a Power Law (PL) as well  as Power Law with an exponential cutoff (PLE) model for a point $\gamma$-ray source at the nominal position of PSR J2051-0827. Photon energies are restricted to 200 MeV to 20 GeV. In the background model, we included the Galactic diffuse model  ({\tt gal\_2yearp7v6\_v0.fits}) and the isotropic background 
({\tt iso\_p7v6source.txt}), as well as all point sources reported in 
the Second Fermi-LAT Source Catalog (2FGL) within $10\degr$ from the center of the ROI. All 
these 2FGL sources were assumed to be point sources which have the specific spectrum suggested in the 2FGL catalog (Abdo et al. 2011). While the spectral parameters of the 2FGL sources locate within the ROI 
were set to be free, we kept the parameters for those lie outside our adopted ROI fixed at the values given in the 2FGL catalog (Abdo et al. 2011). 
We also allowed the normalization factors of the two diffuse background components to vary. The best-fit PL model yields a photon index of $\Gamma=2.46\pm0.15$ and a test-statistic (TS) value of 66 which corresponds 
to a significance of $\sim$8$\sigma$. Its photon flux in this band is found to be $(6.62\pm1.29)\times10^{-9}$~cm$^{-2}$~s$^{-1}$
The corresponding integrated energy flux is $f_{\gamma}=5.92^{+4.75}_{-3.24}\times10^{-12}$~erg~cm$^{-2}$~s$^{-1}$ 
\footnote{The quoted errors
of the energy flux have taken the statistical uncertainties of both photon index and prefactor into account.}. For the PLE model, the best fit results in a photon index of $\Gamma=1.75\pm0.58$ and a cutoff energy of $E_{cutoff}=2.5 \pm 2$~GeV, giving a TS value of 70 which corresponds to a 
significance of $\sim$8.4$\sigma$. Based on the likelihood ratio test, the PLE model are not statistically required to fit the observed 
$\gamma$-ray spectrum. This may be due to the low photon statistic and the contamination from the diffuse $\gamma$-ray background. The binned spectrum is displayed in Figure~\ref{lat_spec}. 

As PSR J2051-0827 is not reported in the 2FGL catalog (Adbo et al. 2011), we tried to reproduce the non-detection result using binned likelihood analysis by defining a $7\degr \times 7\degr$ square ROI, together with the same time span (2 years) and energy cut (0.1-300 GeV) as in the 2FGL catalog (Abdo et al. 2011). The resulting TS is below 25, lower than the selection criteria set by the 2FGL catalog (TS $>$~25). Therefore it is expected that PSR J2051-0827 was not in the 2FGL catalog. After reproducing the non-detection result, we expand the time span by using $\sim$3 years of photon data to perform a  binned likelihood analysis again to check whether PSR J2051-0827 will be detectable with a longer exposure. By assuming a simple power law model, the best-fit model yields a photon index of $\Gamma=2.39 \pm0.13$ and a test-statistic (TS) value of 49. The photon index is consistent with the unbinned likelihood analysis using 2 years of photon data with the same energy cut, and therefore PSR J2051-0827 is detected with $\sim$3-years of LAT data. 

\begin{figure}[t]
\centerline{\psfig{figure=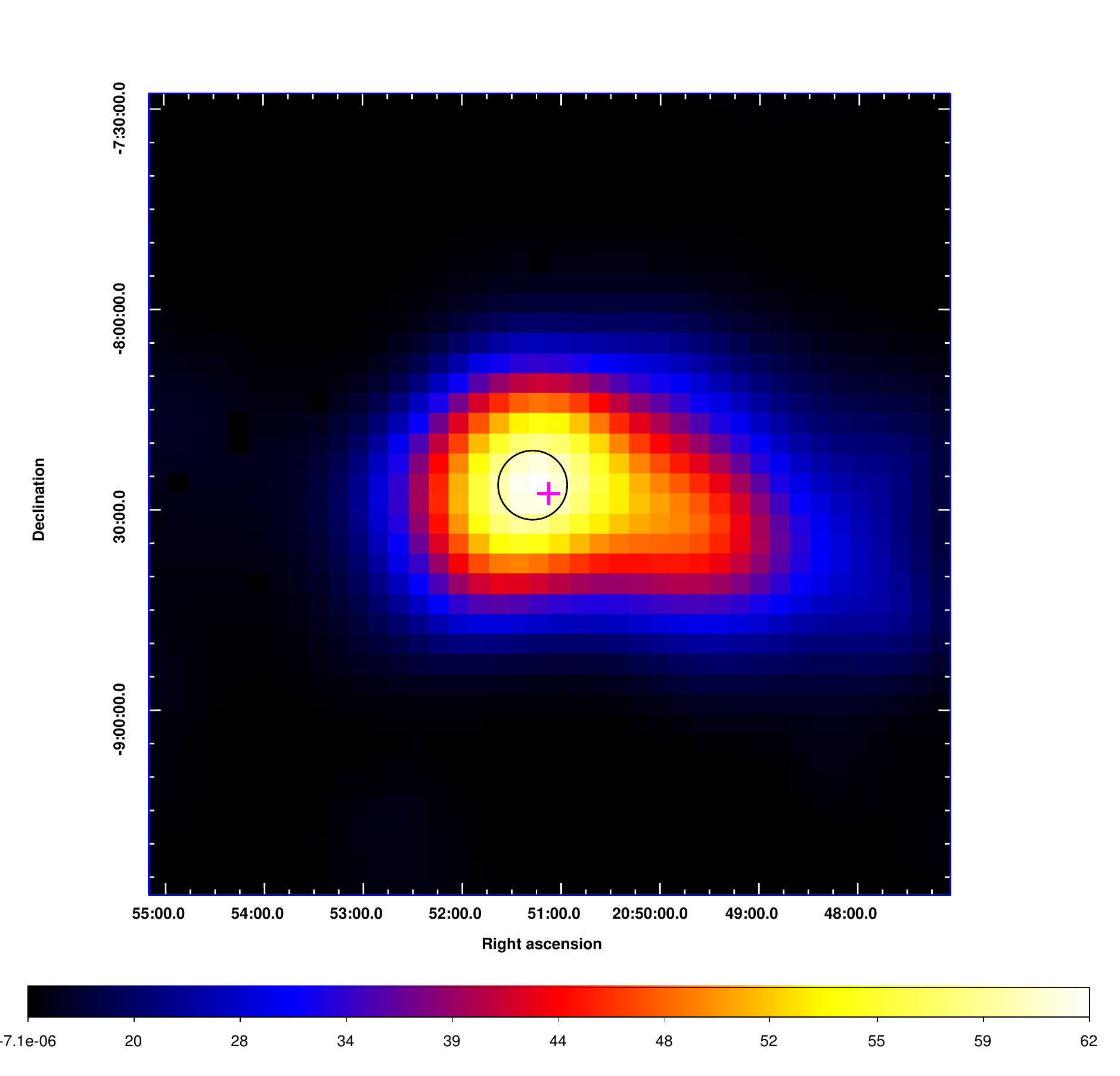,width=8cm,clip=,angle=0}}

\caption{Test-statistic (TS) map in 1-20 GeV of a region of 
2\degr$\times$2\degr~centered at the nominal position of PSR J2051-0827 (magenta cross). 
The color scale that used to indicated the TS values is shown by the scale bar below. 
The circle in black represents the 1$\sigma$ positional error circle determined 
by \textit{gtfindsrc}.}
\label{tsmap_2051}
\end{figure}

We generated a 2\degr $\times$ 2\degr~TS map in 1-20 GeV centered at the nominal 
position of PSR J2051-0827 by using \textit{gttsmap}. This is shown in 
Figure~\ref{tsmap_2051}. With the aid of \textit{gtfindsrc}, 
We determined the best-fit position of PSR J2051-0827 in $1-20$~GeV to be 
RA=20$^{h}$51$^{m}$17.28$^{s}$~ Dec=-08\degr26$^{\arcmin}$22.6$^{\arcsec}$ (J2000) 
with 1$\sigma$~error radius (statistical) of 0.09\degr ~which is illustrated with a black circle 
in Figure~\ref{tsmap_2051}.

\begin{figure}[t]
\centerline{\psfig{figure=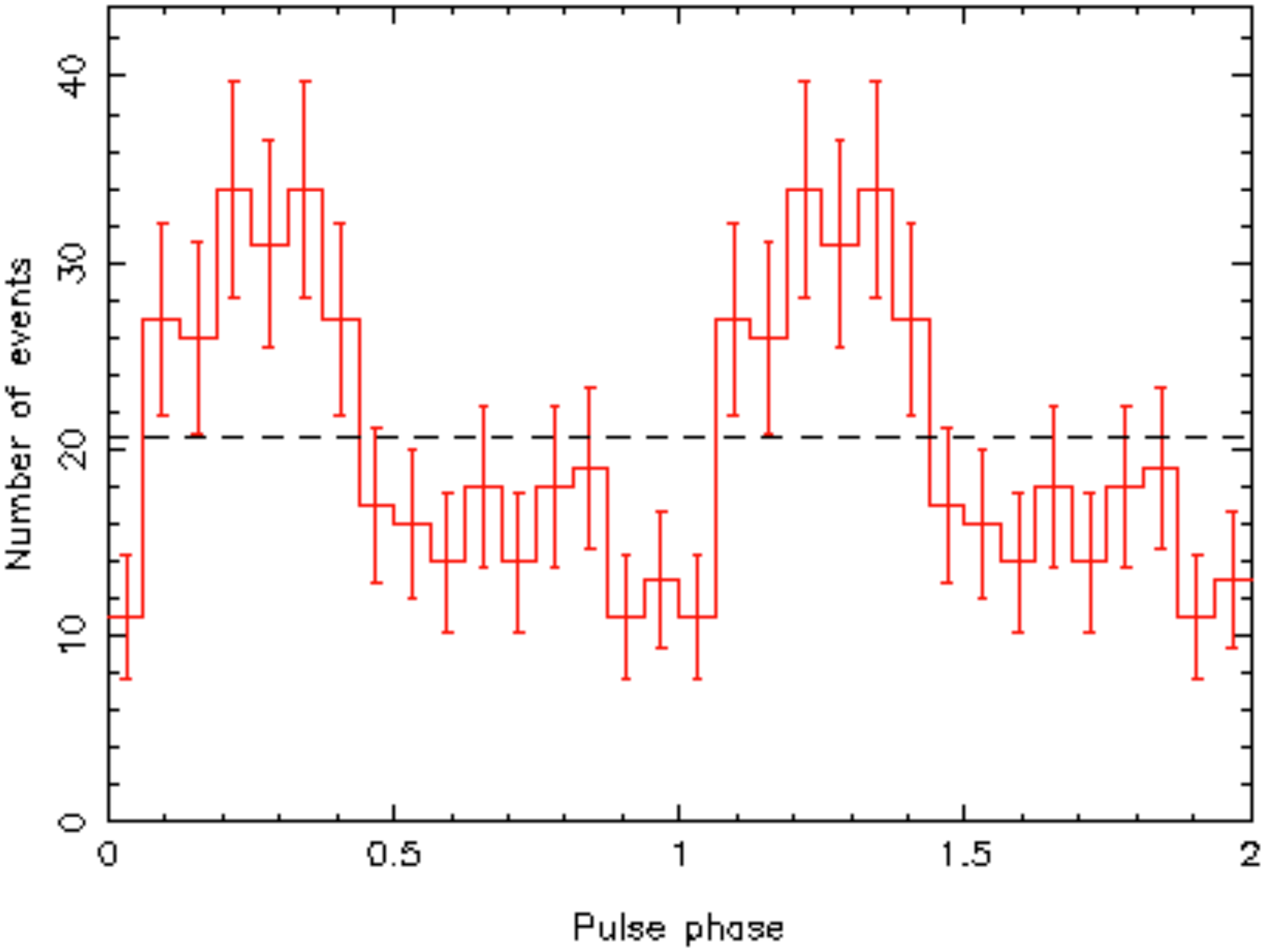,width=8cm,clip=,angle=270}}
\caption{$\gamma$-ray light curve of PSR J2051-0827 using Fermi plugin for TEMPO2 .}
\label{lc_2051}
\end{figure}

By using the latest radio timing ephemeris reported from Lazaridis et al. (2011), we searched for $\gamma$-ray pulsations on PSR J2051-0827 by phase binning {\itshape Fermi} LAT photons received within $\sim$3 years photon data using the Fermi plugin for TEMPO2, {\itshape Fermi} LAT photons were corrected to the solar system barycenter using the JPL DE405 solar system ephemeris. To achieve  the best $\gamma$-ray pulse profile, we test on two parameters: the radius of the aperture ($0.1\degr < r < 1\degr$) and the lowest photon energy (100 $< E_{min} <$ 1000 MeV), with an $E_{max}$ at 300 GeV. The best signal for $\gamma$-ray pulsation of PSR J2051-0827 ($r = 0.5\degr, E_{min} \ge ~$200 MeV) is presented in Figure \ref{lc_2051} with a H-test TS of 35 (de Jager \& B$\ddot{u}$sching 2010), corresponding to a detection significance of 4.9 $\sigma$. We are therefore confident that the newly discovered
$\gamma$-ray source is associated with PSR J2051-0827. 

Since orbital modulations was observed in both radio and optical band, we attempted to search for any evidence on orbital modulations by phase-folding the $\gamma$-ray photons with the orbital parameter using TEMPO2 but no positive result has been found using the orbital parameter obtained by Lazaridis et al. (2011).

\subsection{X-RAY DATA}
PSR J2051-0827 was observed 5 times with the {\it Chandra X-ray
Observatory} on 2009 March 22, 30, and July 5 (ObsID 10106--10110).
All the observations were taken using the Advanced CCD Imaging
Spectrometer array (ACIS-S) with the very faint mode. Data were
collected with a frame transfer time of 3.2 sec and the exposure time
for each observation is about 9 ksec. We used CIAO\footnote{
http://cxc.harvard.edu/ciao/} version 4.3 and
XSPEC\footnote{http://heasarc.nasa.gov/xanadu/xspec/} version 12.7 packages to perform data reduction and analysis. We
reprocessed the raw data to apply the most up-to-date calibration and
to make use the very faint mode. In order to reduce the background, we
restricted the photon energies between 0.3 and 7.0 keV for all our
data analysis. We also inspected the background count rates from the
S1 chip and no flaring event was found in all the observations.

PSR J2051-0827 is marginally seen in each observation with less than
10 counts. We therefore combined all 5 observations with a total
exposure time of $\sim 44$ ksec for subsequent analysis.
We extracted the energy spectrum from a 2 arcsec circular region
centered on PSR J2051-0827; a total of 43 counts were extracted. For
the background, we selected a source-free region with a radius of 30
arcsec, which results 1.1 background count in the
source region. We therefore did not subtract the background for
spectral analysis. Response matrices of each observation were generated and
co-added with CIAO. Because of the low count statistic, we fit the
unbinned spectrum with an absorbed power-law and an absorbed blackbody
models. We fixed the absorption at the Galactic value
($N_H=6\times10^{20}$ cm$^{-2}$) and used CASH statistics (Cash 1979)
to estimate the best-fit parameters and their errors. Both models
provide an acceptable fit. The best-fit photon index is $2.51\pm0.47$
(90\% confidence) yielding an unabsorbed 0.3--10 keV flux of
$8.5\times10^{-15}$ ergs cm$^{-2}$ s$^{-1}$. For the blackbody model,
the best-fit temperature is $kT=0.25\pm0.05$ keV and the unabsorbed
0.3--10 keV flux is $4.9\times10^{-15}$ ergs cm$^{-2}$ s$^{-1}$. We also binned the data with at least 5 counts
per spectral bin and used $\chi^2$ statistics to derive the best-fit
model. The results are consistent with those using CASH statistics.

\section{DISCUSSION}
We report the discovery of the $\gamma$-ray pulsation and the X-ray detection from the second black widow system PSR J2051-0827 using {\itshape Fermi}-LAT and {\itshape Chandra}-ACIS-S data. We compute the rotation energy loss, $\dot{E}$ = $4 \pi^{2} I \dot{P}/P^{3}$, where I is the pulsar's moment of inertia. We assume $I = 10^{45}~g~cm^{2}$ for an ordinary MSP, which gives a $\dot{E}$ = $5.49 \times 10^{33} erg~s^{-1}$. By adopting a distance of 1.04 kpc from Lazaridis et al. (2011), the $\gamma$-ray and the X-ray luminosity are found to be $L_{\gamma}=7.66\times10^{32}erg~s^{-1}$ and $L_{X}=1.01\times10^{30}erg~s^{-1}$ in 0.2-20 GeV and 0.3-10 keV respectively. Therefore the $\gamma$-ray luminosity is $\sim$14\% of the spin down power of PSR J2051-0827, consistent with other MSPs in the Galactic field. 

Comparing PSR J2051-0827 with those black widow pulsars with both $\gamma$-ray
and X-ray emission detected [PSR B1957+20 (Huang \& Becker 2007; Huang et al. in prep.),
PSR J2214+3000 (Ransom et al. 2011) and PSR J2241-5236 (Keith et al.
2011)], they all show similar X-ray properties in terms of their black
body temperature ($\sim 0.25$ keV) and X-ray luminosity ($\sim
10^{30-31}$ erg s$^{-1}$). It is worth noting that except for PSR
B1957+20, all sources have only less than 90 X-ray photons for
spectral analysis and blackbody model may not be the only acceptable
spectral component. For instance, we cannot rule out a power-law model
for PSR J2051-0827. Furthermore, a deep {\it Chandra} observation of
PSR B1957+20 shows that an additional power-law component is required
(Huang et al. in prep.). Unlike the other three black widow systems, PSR
J2051-0827 is a previously uncatalogued $\gamma$-ray source and the
discovery of the $\gamma$-ray emission is entirely driven by the radio
observations. In all four $\gamma$-ray and X-ray emitting black widow pulsars, X-ray pulsation was only detected for PSR B1957+20 (Guillemot
et al. 2011).

Takata et al. (2011) study the X-ray and $\gamma$-ray emissions from the millisecond pulsar in black widow systems. They discuss $\gamma$-ray emission from the outer gap accelerator and  predict  its luminosity to be $L_{\gamma}\sim 8\times 10^{32}~\mathrm{erg s^{-1}}$ for PSR J2051-0827, which is consistent with the results of the present observation.  The observed X-ray emissions from black widow system must be composed of several components. For example, Guillemot et al. (2011) found that the first $\gamma$-ray black widow system, PSR B1957+20, shows that the total X-ray emissions are composed of the non-pulsed ($\sim 67$\%) and pulsed emissions ($\sim 33$\%). The non-pulsed emission will be originated from the synchrotron emissions from the particles accelerated at the intra-binary shock due to the interaction between the pulsar wind and the matter injected from the companion star. For the emissions from the pulsars, the emissions from the heated polar cap region due to the return current or from the magnetospheric synchrotron emissions, or both components can contribute to the observed emissions. 

If the X-ray emissions are from an intra-binary shock, Takata et al. (2011) estimated $L_{X}(0.5-10keV)\sim 2\times 10^{30}$erg/s for PSR J2051-0827, if the magnetization parameters (the ratio of the magnetic energy and the particle energy), is $\sigma \sim 0.1$ and the power index of energy distribution of the emitting particles is $\sim 3$. Such a model will produce an X-ray spectrum with a photon index of $\sim 2$ which is consistent with the observed value. For the magnetospheric synchrotron emissions, Takata et al. (2011) estimate $L_{x}\sim 5\times 10^{29}$erg/s, which may be slightly smaller than the observation.  For 0.3-10~keV bands, so called ÒcoreÓ component of the heated polar cap emissions, which has a temperature $\sim 10^6$~K and the effective radius $~10^4$cm, will contribute to the observed emissions. Using equation (28) in Takata et al. (2011), the temperature and luminosity of the core component for PSR J2051-0827 are estimated as $\sim 0.25$~keV and  $L_{x}\sim 4\times 10^{30}~\mathrm{erg/s}$, respectively, where the effective radius is assumed  to be $R\sim 0.1$~km. This is consistent with the observed values.    

Note that Lazaridis et al. (2011) have shown the secular variations on orbital period on PSR J2051-0827, in which the timing model is valid until MJD 54888, which do not cover the whole range of the \textit{Fermi}-LAT data used in folding the $\gamma$-ray light curve. Although there is no indication of any orbital period variation in the $\gamma$-ray light curve but as the orbital parameter are subjected to change over time, further radio/optical observations are necessary to refine the $\gamma$-ray pulse profile.

This project is supported by the National Science Council of the
Republic of China (Taiwan) through grant NSC100-2628-M-007-002-MY3 and
NSC100-2923-M-007-001-MY3. A.K.H.K. gratefully acknowledges support
from a Kenda Foundation Golden Jade Fellowship. J.T. and K.S.C. are supported by a GRF grant of HK Government under HKU700911P.

\end{document}